\documentclass[preprint2]{aastex631}

\accepted{July 28, 2022}

\submitjournal{PASP}

\shorttitle{Performance Assessment of the K-DRIFT pathfinder}
\shortauthors{Byun et al.}
\graphicspath{{./}{figures/}}
\begin{document}

\title{Performance Assessment of the KASI-Deep Rolling Imaging Fast-optics Telescope pathfinder}

\correspondingauthor{Woowon Byun, Jongwan Ko}
\email{wbyun87@gmail.com, jwko@kasi.re.kr}

\author[0000-0002-7762-7712]{Woowon Byun}
\affiliation{Korea Astronomy and Space Science Institute, Daejeon 34055, Republic of Korea}

\author[0000-0002-9434-5936]{Jongwan Ko}
\affiliation{Korea Astronomy and Space Science Institute, Daejeon 34055, Republic of Korea}
\affiliation{University of Science and Technology, Korea, Daejeon 34113, Republic of Korea}

\author[0000-0003-0009-5161]{Yunjong Kim}
\affiliation{Korea Astronomy and Space Science Institute, Daejeon 34055, Republic of Korea}

\author[0000-0001-9561-8134]{Kwang-Il Seon}
\affiliation{Korea Astronomy and Space Science Institute, Daejeon 34055, Republic of Korea}
\affiliation{University of Science and Technology, Korea, Daejeon 34113, Republic of Korea}

\author{Seunghyuk Chang}
\affiliation{Center for Integrated Smart Sensors, Daejeon 34141, Republic of Korea}

\author{Dohoon Kim}
\affiliation{Green Optics Co., Ltd., Cheongju 28126, Republic of Korea}

\author{Changsu Choi}
\affiliation{Korea Astronomy and Space Science Institute, Daejeon 34055, Republic of Korea}

\author[0000-0002-6154-7558]{Sang-Hyun Chun}
\affiliation{Korea Astronomy and Space Science Institute, Daejeon 34055, Republic of Korea}

\author{Young-Beom Jeon}
\affiliation{Korea Astronomy and Space Science Institute, Daejeon 34055, Republic of Korea}

\author[0000-0002-1710-4442]{Jae-Woo Kim}
\affiliation{Korea Astronomy and Space Science Institute, Daejeon 34055, Republic of Korea}

\author[0000-0003-0043-3925]{Chung-Uk Lee}
\affiliation{Korea Astronomy and Space Science Institute, Daejeon 34055, Republic of Korea}

\author[0000-0001-7594-8072]{Yongseok Lee}
\affiliation{Korea Astronomy and Space Science Institute, Daejeon 34055, Republic of Korea}
\affiliation{School of Space Research, Kyung Hee University, Yongin, Kyeonggi 17104, Republic of Korea}

\author[0000-0002-3505-3036]{Hong Soo Park}
\affiliation{Korea Astronomy and Space Science Institute, Daejeon 34055, Republic of Korea}
\affiliation{University of Science and Technology, Korea, Daejeon 34113, Republic of Korea}

\author{Eon-Chang Sung}
\affiliation{Korea Astronomy and Space Science Institute, Daejeon 34055, Republic of Korea}

\author[0000-0002-6841-8329]{Jaewon Yoo}
\affiliation{Korea Astronomy and Space Science Institute, Daejeon 34055, Republic of Korea}
\affiliation{University of Science and Technology, Korea, Daejeon 34113, Republic of Korea}

\author{Gayoung Lee}
\affiliation{Kyungpook National University, Daegu 41566, Republic of Korea}

\author{Hyoungkwon Lee}
\affiliation{LeO SPACE Inc., Daejeon 34014, Republic of Korea}

%% Note that the \and command from previous versions of AASTeX are now
%% automatically takes care of all commas and "and"s between authors' names.


\begin{abstract}

In a $\Lambda$CDM universe, most galaxies evolve by mergers and accretions, leaving faint and/or diffuse structures, such as tidal streams and stellar halos. Although these structures are a good indicator of galaxies' recent mass assembly history, they have the disadvantage of being difficult to observe due to their low surface brightness (LSB). To recover these LSB features by minimizing the photometric uncertainties introduced by the optical system, we developed a new optimized telescope named K-DRIFT pathfinder, adopting a linear astigmatism free-three mirror system. Thanks to the off-axis design, it is expected to avoid the loss and scattering of light on the optical path within the telescope. To assess the performance of this prototype telescope, we investigate the photometric depth and capability to identify LSB features. We find that the surface brightness limit reaches down to $\mu_{r,1\sigma}\sim28.5$ mag arcsec$^{-2}$ in $10^{\prime\prime}\times10^{\prime\prime}$ boxes, enabling us to identify a single stellar stream to the east of NGC 5907. We also examine the characteristics of the point spread function (PSF) and find that the PSF wing reaches a very low level. Still, however, some internal reflections appear within a radius of $\sim$6 arcmin from the center of sources. Despite a relatively small aperture (0.3 m) and short integration time (2 hr), this result demonstrates that our telescope is highly efficient in LSB detection. 

\end{abstract}

%% The AAS Journals now use Unified Astronomy Thesaurus concepts:
%% You will be asked to select these concepts during the submission process
\keywords{Multiple mirror telescopes (1080), Astronomy data reduction (1861), Galaxies (573), Stellar streams (2166)}

%% and \citet commands to identify citations. The citations are

\section{Introduction} \label{sec:intro}

In the $\Lambda$CDM model, galaxies undergo numerous mergers and accretions \citep{1978MNRAS.183..341W}. Small galaxies orbiting large ones are occasionally disrupted by tidal forces, leaving them as stellar streams \citep[see][]{2001ApJ...548...33B,2005ApJ...635..931B}. These structures gradually disperse for a few Gyr and eventually become a part of stellar halos of host galaxies \citep[cf.][]{2008MNRAS.391...14D,2008ApJ...689..936J}. That is, most large galaxies in the universe are expected to have extended, faint tidal remnants \citep[e.g.,][]{1999ApJ...524L..19M,2010MNRAS.406..744C,2013MNRAS.434.3348C}. In other words, we can directly infer the recent assembly history of a host galaxy via the existence of stellar streams and stellar halos and the nature of progenitors via their shapes, colors, and total luminosities. Such a description can be applied to not only the structures in the outskirts of individual galaxies but also the intracluster light in cluster environments \citep[e.g.,][]{2007ApJ...668..826C,2014MNRAS.437.3787C}. 

To verify this paradigm, decisive observational evidence is necessary. Unfortunately, however, there is a critical limitation that it is difficult to observe such structures because of their low stellar surface density, which generally corresponds to a surface brightness of at least $\sim$27--28 mag arcsec$^{-2}$ \citep[cf.][]{2011ApJ...739...20C,2012arXiv1204.3082B}. 

Thanks to its relative closeness, many observational studies have been dedicated to the Local Group, including the Milky Way and Andromeda, using the star count method \citep[e.g.,][]{1997AJ....113..634I,2007ApJ...658..337B,2011ApJS..195...18R,2014ApJ...787...19M,2018ApJ...868...55M}. The photometric depth achieved by this method corresponds to at least $\sim$30 mag arcsec$^{-2}$. To acquire more samples beyond the Local Group, the measurement of smooth integrated light of galaxies has been carried out as an alternative approach \citep[e.g.,][]{1997ASPC..116..460M,1998ApJ...504L..23S,2005ApJ...631L..41M,2005AJ....130.2647V,2009AJ....138.1417T,2010AJ....140..962M,2013ApJ...765...28A,2018ApJ...862...95K,2021MNRAS.508.2634Y}. Still, it appeared to barely achieve surface brightness levels of $\sim$28--29 mag arcsec$^{-2}$. Thus, overcoming the existing observational limitation is essential to investigating the nature of extremely low surface brightness (LSB) galaxies. 

The reason why LSB imaging is difficult is not just because of insufficient exposure time. Systematic uncertainties are more likely to be the primary cause of this difficulty. For example, reaching deeper surface brightness levels will be hindered by local background fluctuations, introduced by a combination of zodiacal light, moonlight, stray light from stars, point spread function (PSF), and the Galactic cirrus, etc. \citep[see][]{2017ASSL..434.....K,10.1088/2514-3433/ac2c7d}. Those factors can cause inaccuracy in the calibration process, such as sky determination, resulting in significant photometric uncertainty in the LSB regime. In addition, they can directly interfere with LSB detection. For example, spurious light derived from local background fluctuations may be mistaken for LSB features. 

Specific observation strategies, e.g., observing objects with high galactic altitudes in a dark time, can potentially reduce fluctuations caused by external factors. The remaining problem is that ordinary telescopes may hold a risk of the loss and/or scattering of light on the optical path. Large diameters are indeed advantageous for collecting lots of photons. However, larger telescopes are likely to have more complex optical systems, which can cause various problems. For example, a significant PSF wing may affect the background, thereby obscuring LSB features or leading to incorrect estimates of the LSB galaxy's properties such as shape and brightness \citep[e.g.,][]{2016ApJ...823..123T}. In addition, internal reflection can create extended asymmetric artifacts around bright stars, commonly referred to as \textit{ghosts}, which significantly contaminate images. Indeed, we may fail to detect the LSB objects if they are overlapped with the artifacts brighter than 27 mag arcsec$^{-2}$. 

While software processing through ray tracing and machine learning can be applied to remove these artifacts \citep[e.g.,][]{2009PASP..121.1267S,2022A&C....3900580T}, we still need to develop telescopes optimized for LSB studies that suppress loss and scattering of light fundamentally, through the careful baffling, anti-reflective coating, as well as simple optics design. The Dragonfly Telephoto Array made with multiple commercial CCD cameras is a good example of a novel attempt to overcome surface brightness limitation \citep[see][]{2014PASP..126...55A}. In brief, it was designed to be able to reach a deep surface brightness level and obtain a wide field of view (FoV) at the expense of a high spatial resolution. 

As another independent attempt to do so, we launched a project at the Korea Astronomy and Space Science Institute (KASI) in 2019. It is called ``KASI-Deep Rolling Imaging Fast-optics Telescope (K-DRIFT)''. This project is planning to develop new telescopes optimized for LSB studies and conduct deep imaging surveys. We adopted an off-axis design with a small focal ratio to improve the response to faint light, even with a small aperture. As a prototype telescope, the K-DRIFT pathfinder is significant because it is the first ground-based telescope manufactured with this concept. Over the next decade, we plan to build several 0.3--0.5 m ground-based telescopes and even launch a small space telescope (see details in Ko et al. 2022 in preparation).

%A telescope with a small focal ratio is advantageous for observing faint objects even if the aperture size is small. Taking advantage of this characteristic, the project is planning to manufacture several ground-based telescopes with a diameter of 0.3-0.5m and a small space telescope over the next ten years (see details in Ko et al. in preparation). 

The manufacture of the K-DRIFT pathfinder was completed in early 2021, and we installed it at the Bohyunsan Optical Astronomy Observatory (BOAO) in Korea for test observation. In this paper, we assess the performance of the K-DRIFT pathfinder and discuss the potential of the project by understanding the systems that must be improved in the next production of telescopes. The paper is organized as follows: Section \ref{sec:telescope} describes the specification of the telescope, and Section \ref{sec:data} presents an outline of the observation and data reduction. Section \ref{sec:result} provides the observational results and describes the telescope's performance. Section \ref{sec:disc} discusses some shortcomings of the telescope, and Section \ref{sec:sum} summarizes all the assessment.

\section{Telescope} \label{sec:telescope}

%% The "ht!" tells LaTeX to put the figure "here" first, at the "top" next
%% and to override the normal way of calculating a float position
\begin{figure}[t]
\centering
\vspace{3mm}
\includegraphics[width=80mm]{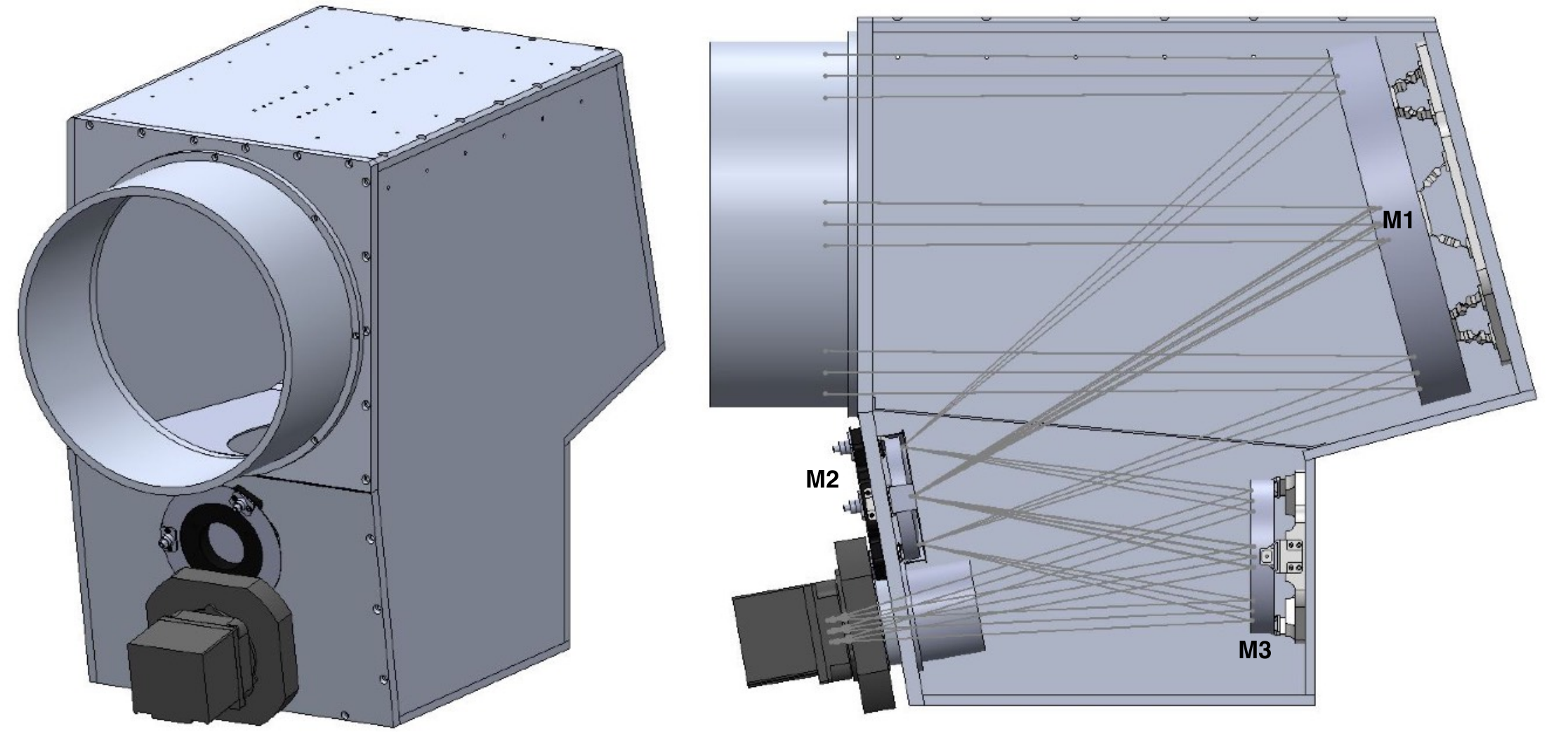}
\caption{Exterior and interior designs of the K-DRIFT pathfinder. The diameter of the aperture is 300 mm. \label{fig:fig1}}
\end{figure}

Figure \ref{fig:fig1} shows the configuration of the K-DRIFT pathfinder. It was designed as a linear astigmatism free-three mirror system (LAF-TMS) that is an all-reflective imaging system consisting of three free-form mirrors \citep[cf.][]{10.1117/12.2023433,2020PASP..132d4504P}. The off-axis design can minimize the loss and scattering of light due to not requiring the internal structures located on the optical path. In addition, it is analytically optimized to remove linear astigmatism and reduce aberrations. Thus, it can produce a small and nearly-uniform spot size across the entire FoV, thereby avoiding quality degradation for the wide-field image. Specifically, our system was designed to have an average spot size of 3.6 $\mu$m, corresponding to 0.3 pixels. The mirrors are composed of a concave primary mirror (M1), a convex secondary mirror (M2), and a concave tertiary mirror (M3). All the mirrors were coated with silver to enhance the reflectivity. Triple baffling can also robustly remove stray light from stars outside the line of sight.

The diameter of the aperture is 300 mm, and the focal length is 1200 mm, resulting in the focal ratio of $f/4$. A 2k$\times$2k sCMOS camera (KL400 BI) is installed and the FoV corresponds to $\sim$$1\times1$ deg$^2$, yielding a pixel scale of $\sim$1.89 arcsec. Only a clear luminance ($L$) filter, which covers a range of 3500--8500 $\mathrm{\AA}$, is equipped to collect as many photons as possible. The CMOS camera's quantum efficiency has been improved significantly compared to the past, resulting in more than 90\% at the effective wavelength. 

Note that CCD cameras, widely used in astronomy, have characteristics that readout time inevitably increases as the FoV and angular resolution increase. Dividing a CCD chip into several amplifiers is often used to resolve this issue. However, this can introduce artificial patterns and uncertainty when flattening images, thus hampering the detection of LSB signals. In this regard, CMOS has the advantage of being able to operate relatively fast, even for a single large chip.\footnote{With the telescope's specifications mentioned above, readout time may not be a significant problem even if a CCD camera was used. However, we had to consider that the next K-DRIFT telescopes will adopt imaging sensors with a resolution of at least 6k $\times$ 6k.} This advantage allows the development of an optimized observation strategy, including frequent and rolling dithering, to process imaging data more precisely. Also, the small pixels of CMOS sensors make it possible to reduce the optical system's size, power consumption, and cost. However, there are still shortcomings, such as high dark current and poor electronic stability, but we expect it will improve rapidly soon. Despite some issues, the combination of novel telescopes and CMOS is expected to secure originality from previous studies. 

It is worth mentioning that this is the first time the whole development process of a LAF-TMS was completed and did not end up as a laboratory experiment but led to actual observations. By doing so, we could clarify the advantages and disadvantages of the telescope. These assessments will be utilized in the upcoming development of the next K-DRIFT telescopes. The details about the fabrication, assembly, and alignment process can be found in Y. Kim et al. (2022 in preparation). 

\section{Data}\label{sec:data}

\subsection{Observation} \label{sec:data:obs}

The observation to verify the telescope's performance was conducted on 2021 June 5 and 6 at BOAO in Korea. As the first target to observe LSB structures, we selected an edge-on spiral galaxy, NGC 5907. This galaxy, which has a stellar mass of $\sim$$8\times10^{10}\ M_\odot$ \citep{2016AJ....152...72L}, is one of the well-known galaxies with giant stellar stream(s) in the Local Group ($D\sim 17$ Mpc; \citealt{2011ApJS..195...18R}). The stream was first reported by \cite{1998ApJ...504L..23S}, and many observational verifications and analyses have been carried out until recently \citep[e.g.,][]{1999AJ....117.2757Z,2008ApJ...689..184M,2016AJ....152...72L,muller+2019,2019ApJ...883L..32V}. This allows us to compare our results with other well-studied results and effectively assess the performance of the K-DRIFT pathfinder. In addition, NGC 5907 and its vicinity are likely to be less contaminated by the Galactic cirrus since the far-infrared flux appears to be low ($F_\mathrm{100\mu m}\lesssim 1.5$ MJy sr$^{-1}$) from the IRAS map.\footnote{\url{https://irsa.ipac.caltech.edu}} Thus, the following results are likely to mainly reflect the telescope's performance without being significantly affected by external factors. 

%% The "ht!" tells LaTeX to put the figure "here" first, at the "top" next
%% and to override the normal way of calculating a float position
\begin{figure}[t]
\centering
\vspace{3mm}
\includegraphics[width=80mm]{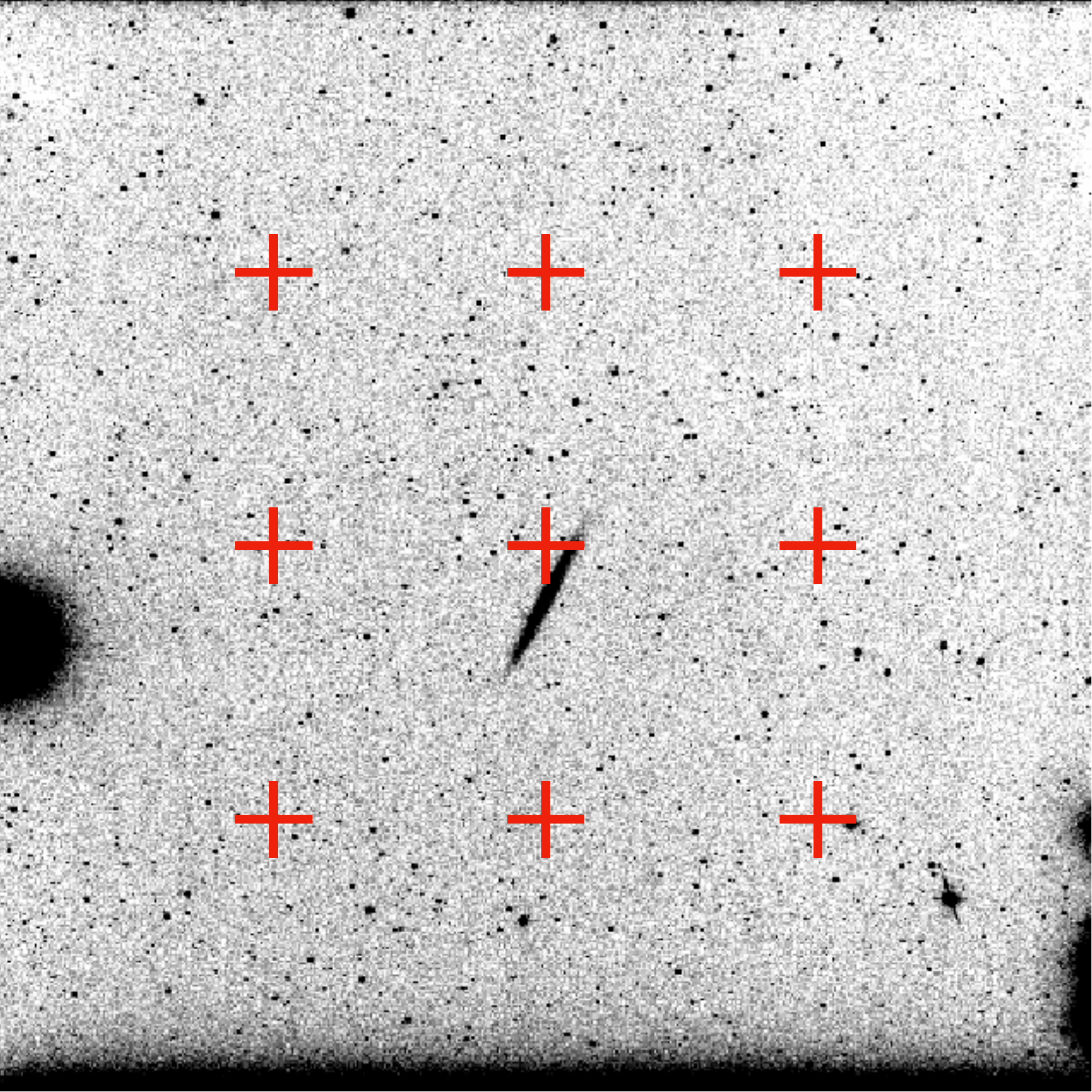}
\caption{Schematic dithering pattern with red crosses plotted over a raw image of NGC 5907 taken by the K-DRIFT pathfinder. The FoV of the image corresponds to $\sim$$1\times1$ deg$^2$. The gap between adjacent dithering positions is $\sim$15 arcmin. Amplifier glow is distinct at the image boundary. \label{fig:fig2}}
\end{figure}

Figure \ref{fig:fig2} shows a raw image of NGC 5907 together with the schematic dithering pattern we used. We observed NGC 5907 by positioning it at nine points during one night. Since the diameter of NGC 5907 corresponds to $\sim$ 13 arcmin \citep{1973UGC...C...0000N}, we chose the gap of nine points to be at least $\sim$15 arcmin so that the structure of NGC 5907 is not overlapped between every dithers. We performed a non-constant shift so that each position has a slight offset in practice. A 30-sec exposure was taken ten times continuously at each place. Due to the low readout noise of the CMOS camera (1.6 electrons), each frame would be sky-noise limited in short exposure times. This ensures a large number of dithered object frames in a relatively short period, which can be advantageous for minimizing the background noise in the final co-added image. As a result of two-day iterative observations, the total exposure time is 2 hr. 

As described in the next section, we performed flat fielding using a dark-sky flat generated by combining object-masked science images. To do so, considerable dithering was necessary. Such dithering provides enough pixels of the blank sky to generate a dark-sky flat robustly. It can also remove artifacts, such as cosmic rays and satellite tracks, when co-adding images in the final data processing. 

%% The "ht!" tells LaTeX to put the figure "here" first, at the "top" next
%% and to override the normal way of calculating a float position
\begin{figure}[t]
\centering
\vspace{3mm}
\includegraphics[width=80mm]{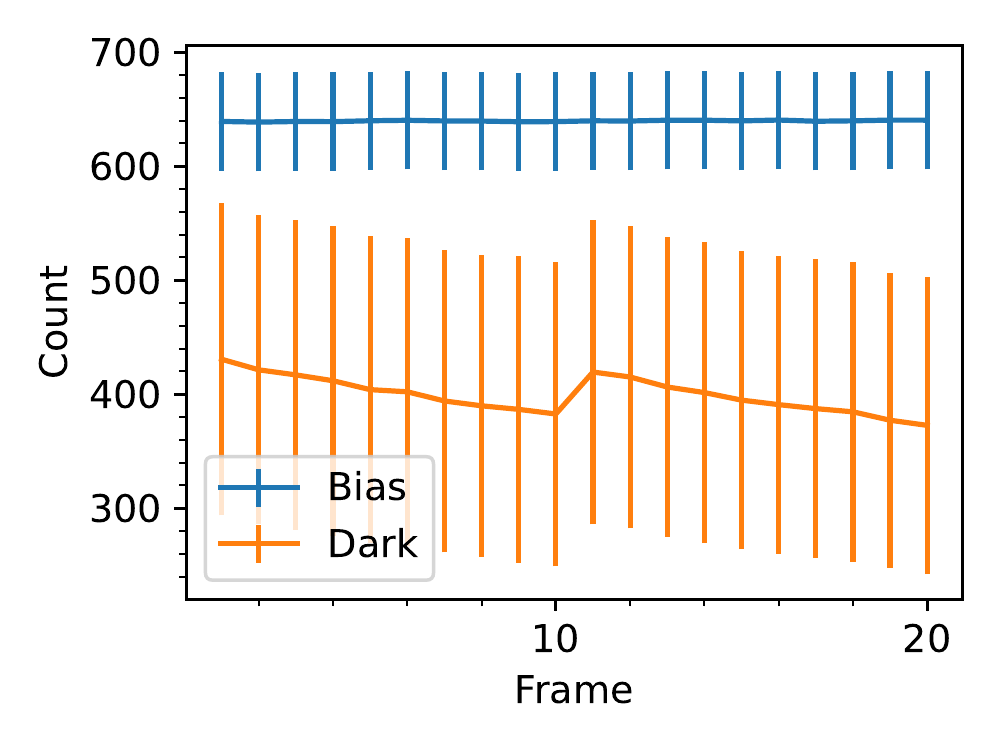}
\caption{Variation in the levels of bias (blue) and bias-subtracted dark (orange). The error bars indicate the standard deviation of each frame. Frames ``1-10'' and ``11-20'' were continuously obtained at once after target observation. The bias level appears nearly constant, but the dark level shows a decreasing trend over time. \label{fig:fig3}}
\end{figure}

\subsection{data reduction} \label{sec:data:reduc}

The following procedures were performed using a custom-developed pipeline based on PYTHON, including \texttt{astropy} and \texttt{photutils} libraries. This pipeline is expected to be flexible in dealing with data of the next K-DRIFT telescopes, which will be developed for a decade. The overall concept of data reduction was adopted from \cite{2018AJ....156..249B}. 

Before data processing, we checked the variation of bias and dark levels to ensure the reliability of master calibration frames. Figure \ref{fig:fig3} shows the median values of bias and bias-subtracted dark frames. Here, frames ``1--10'' and ``11--20'' were independently obtained after different target observations. In fact, we carried out object and calibration observations alternately to examine the detector's response. The bias level appears almost constant with a slight deviation, while the dark level seems pretty complicated. 

The dark level, $\sim$420--430 ADU, steadily decreased over 300 sec and reached down to $\sim$370--380 ADU. Since the detector has been cooled down to approximately $-20^\circ\mathrm{C}$, we ruled out the possibility of contamination from external heat sources. At the same time, we found that the dark frames had false signals where the bright stars were placed on the last object frame, and these signals diminished over time. It is very similar to the phenomenon known as ``residual bulk image (RBI)''. Thus, we suspect that the dark (object) frames seem to be affected by the object image taken just before, which has a background level of $\sim$13000 ADU on average. This may be a critical issue in that the detector cannot completely flush the charge of each exposure. We will discuss it again in Section \ref{sec:disc}. 

Meanwhile, the peak-to-peak deviation of the dark level was revealed as $\sim$0.5\% of the background level of object images. Therefore, it should be considered that using a master dark can lead to a non-negligible uncertainty in dark subtraction. Fortunately, however, the deviation is not too large, and each dark frame seems not to have any distinct pattern over the images. Thus, this may not be a critical flaw in identifying the existence of any LSB structures. However, there is still a potential error in measuring surface brightness. 

%% The "ht!" tells LaTeX to put the figure "here" first, at the "top" next
%% and to override the normal way of calculating a float position
\begin{figure}[t]
\centering
\vspace{3mm}
\includegraphics[width=80mm]{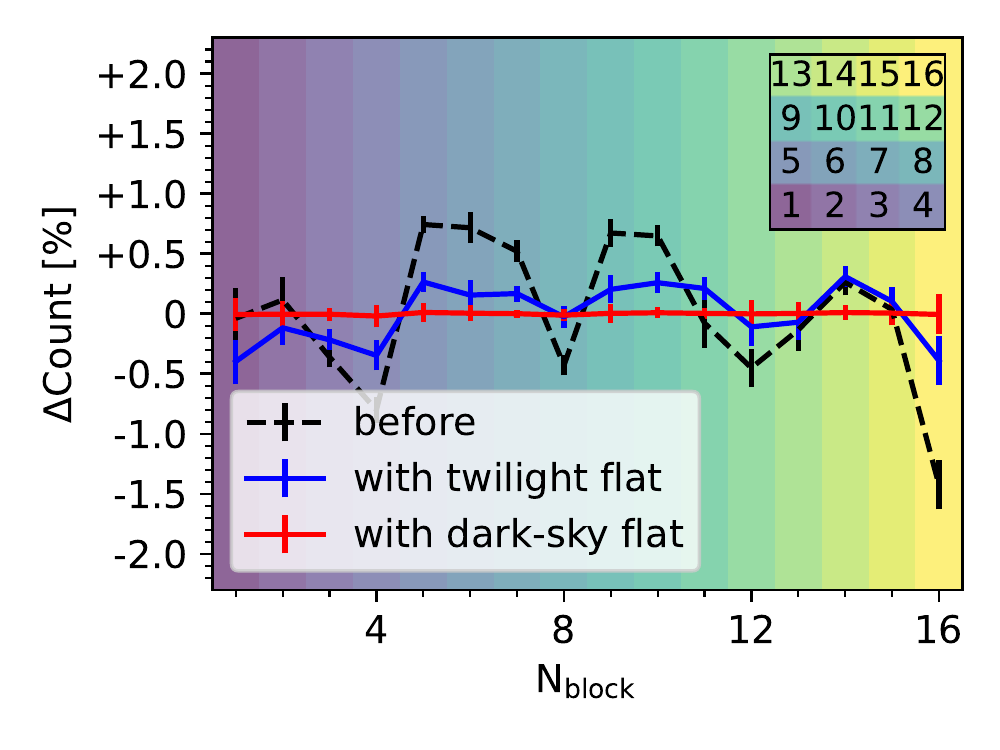}
\caption{Comparison of background variations of images divided into 16 blocks using a $4\times4$ grid. The dashed black line represents the median value of randomly-selected 90 images before flat fielding. The solid blue and red lines represent the median values of the same frames but after flat fielding with twilight flat and dark-sky flat, respectively. The error bars indicate the standard deviation between 90 images. The block numbers on the x-axis follow the order row-by-row, bottom-to-top, and each block is color-coded to clarify the order. \label{fig:fig4}}
\end{figure}

After bias and dark subtraction, flat fielding was carried out using a dark-sky flat. To create a dark-sky flat, we first masked all bright objects in the science images with a 2$\sigma$ threshold of the background noise. The mask size was then slightly expanded by convolution using a Gaussian kernel. In addition, we manually added circular masks for bright stars and galaxies with a radius twice the major axis of each mask to cover unmasked diffuse light. Finally, the normalized object-masked science images were median-combined. The data may have different photometric conditions between two other nights, so we created two dark-sky flats using only the images obtained on the same night and separately used them for flat fielding.

Figure \ref{fig:fig4} shows the results of flat fielding for a subsample of randomly-selected 90 images. To visualize the background variation effectively, we divided each image into 16 blocks using a $4\times4$ grid. We found a gradient from left to right in the images before the flat fielding, and the peak-to-peak deviation among 16 blocks was revealed as $\sim$2\% of the original background level (dashed black line). On the other hand, the sky gradient became negligible after the flat fielding with the dark-sky flat, and the deviation decreased to as low as $\sim$0.02\% (solid red line). 

It is almost impossible to secure a flat, wide light source surface. Therefore, obtaining the actual pixel-to-pixel sensitivity variation in a wide-field image is very difficult. Using a dark-sky flat can be a good approach, but there is a risk of eliminating the real sky gradients, mainly introduced by the variation of airglow depending on the zenith angle. Instead, we expect that the `median-combining' of a sufficiently large number of frames provides a master flat almost unrelated to each frame's real gradient. Indeed, the gradients of all the object frames used to create the dark-sky flat varied even from frame to frame. The deviation was less than 0.3\%, but the trend appeared nearly consistent. 

On the other hand, twilight flat frames may significantly differ in trend, as they are literally obtained at different times and/or from different skies. To verify whether the twilight flat is suitable for flattening images, we carried out flat fielding with the twilight flat. A total of 45 twilight flat frames were carefully taken at zenith with a slight offset. As shown in Figure \ref{fig:fig4} (solid blue line), there was still a significant fluctuation of 1\% level. This fluctuation might be removed by subtracting the background using complex sky models in the following procedure. However, this approach can cause another photometric uncertainty. In addition, unlike the dark-sky flat, the twilight flat could not effectively remove the amplifier glow because of different exposure times. Thus, using a dark-sky flat is likely to be more robust in flattening wide-field images than using a twilight flat. 

We performed sky subtraction in the individual images. Although the flat fielding almost mitigated the sky fluctuations, two-dimensional background models were used to eliminate the remaining trivial sky gradients. Briefly, we divided an object-masked image into a $16\times16$ grid and measured the median values of each. Then, we generated a pixel-scaled background model with a second-order two-dimensional polynomial fitting. Using a smaller or larger grid may change the background model, but it did not significantly affect the result of this study. 

The world coordinate system of the resulting images was updated using \texttt{astrometry.net},\footnote{\url{https://astrometry.net}} and then all images were co-added using SWarp \citep{2002ASPC..281..228B} with median-combine. Note that the dark subtraction removed most of the amplifier glow but left a small spot on the left side of the images. For that reason, we masked that area in all images before co-addition so that it does not affect the result at all. 

%% The "ht!" tells LaTeX to put the figure "here" first, at the "top" next
%% and to override the normal way of calculating a float position
\begin{figure}[t]
\centering
\vspace{3mm}
\includegraphics[width=80mm]{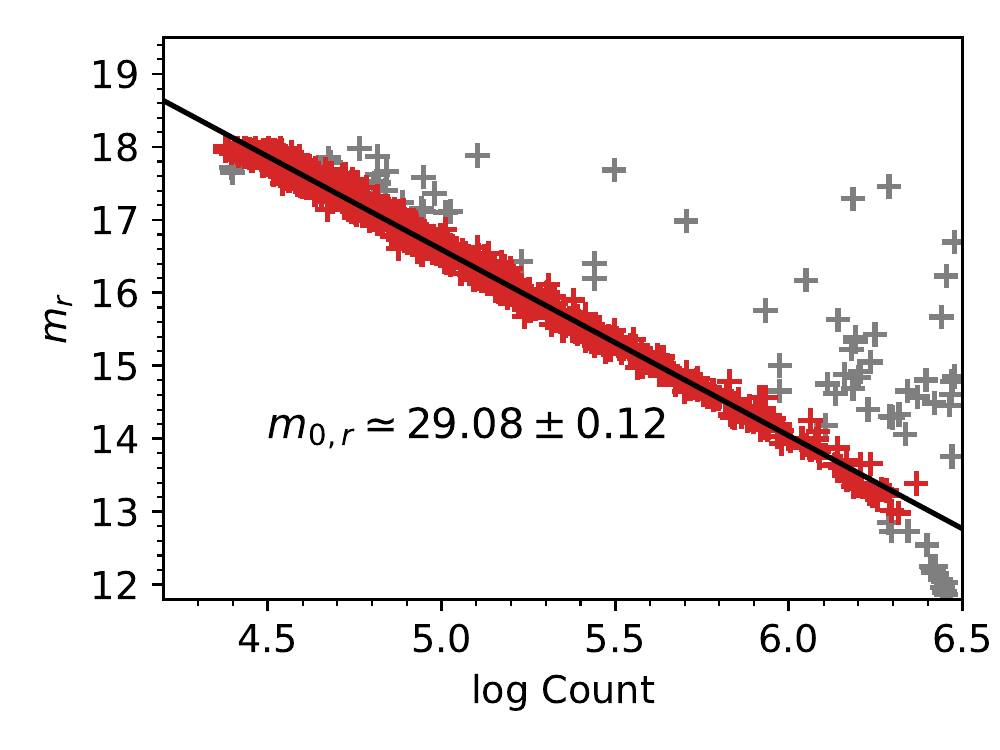}
\caption{SDSS $r$ magnitudes of the field stars in the final co-added image as a function of their total counts estimated using aperture photometry. The contaminated or saturated ones (gray) were excluded by iterative sigma clipping, and the rest (red) were used for a linear fit (solid line). \label{fig:fig5}}
\end{figure}

Standardization was performed using the Sloan Digital Sky Survey (SDSS) $r$-filter data since no archive data is available in the $L$-band. We utilized the magnitudes of $\sim$870 field stars within a radius of 30 arcmin from the center of NGC 5907 and measured the fluxes of the stars in units of counts using \texttt{FLUX\_BEST} in SExtractor \citep{1996A&AS..117..393B}. As shown in Figure \ref{fig:fig5}, we estimated the zero point by using $\sim$90\% of the matched stars, excluding defects and saturated ones. It appears that stellar magnitudes versus counts are nicely fitted by linear regression, resulting in the zero point of $m_{0,r}\simeq29.08\pm0.12$. This nominal value can go down one mag deeper if the $g$ filter is applied instead. Here, one can see that the luminance filter has advantageous for LSB \textit{detection}, but there is a limitation in quantitative comparison with data obtained using more common filters. Therefore, we are considering the application of various filter sets for detecting and analyzing LSB features in the future. 

\section{Results}\label{sec:result}

\subsection{Identification of LSB features}

We remind the readers that one of the main goals of this project is to detect LSB features such as tidal streams in the outskirts of galaxies, (diffuse) dwarf galaxies in the nearby universe, and intracluster light of galaxy clusters. The typical size of these sources is expected to be considerably larger than the pixel size of our data. Therefore, correction for their actual angular sizes must be applied to estimate the practical detection limit. 

%% The "ht!" tells LaTeX to put the figure "here" first, at the "top" next
%% and to override the normal way of calculating a float position
\begin{figure*}[t]
\centering
\vspace{3mm}
\includegraphics[width=180mm]{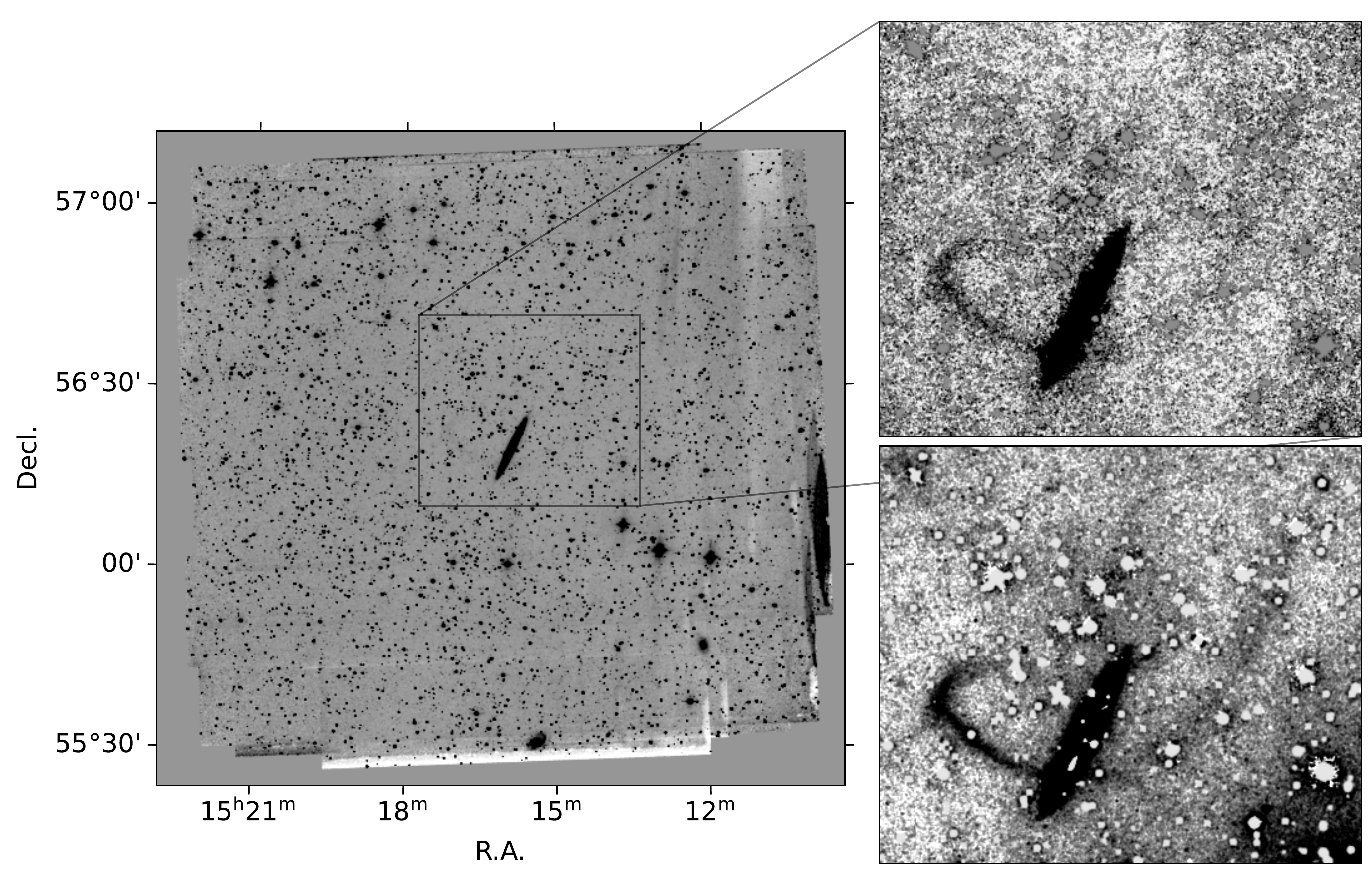}
\caption{Left: the final co-added image of NGC 5907 field. Right: the zoomed-in image of NGC 5907 in ours (top) and the $g$-filter image provided from \cite{2019ApJ...883L..32V} (bottom). The FoV of the two is the same, and all bright objects except NGC 5907 were masked. Both images were smoothed with a Gaussian filter to emphasize LSB structures. A stellar stream to the east of the galaxy is evident in both, but our result does not clearly identify a long tail to the northwest. \label{fig:fig6}}
\end{figure*}

Referring to the definition of \cite{2020A&A...644A..42R}, we estimated the 1$\sigma$ surface brightness limit from the background noise, yielding $\mu_{r,1\sigma}\sim28.5$ mag arcsec$^{-2}$ in $10^{\prime\prime}\times10^{\prime\prime}$ boxes. Note that the surface brightness limit can vary depending on ways of measuring surface brightness. For instance, a one-dimensional radial surface brightness profile of galaxies can be used to detect much fainter structures. Therefore, the above nominal value may be considered an upper limit to our sensitivity. 

Figure \ref{fig:fig6} shows the final co-added and zoomed-in images of NGC 5907, which directly demonstrate the telescope's performance in detecting LSB features. To emphasize the LSB features in the zoomed-in image, all sources but those related to NGC 5907 were masked, and the image was smoothed with a Gaussian function. In short, we were able to clearly identify a single curved stellar stream to the east of NGC 5907, as already reported in the literature \citep[e.g.,][]{1998ApJ...504L..23S,2016AJ....152...72L,muller+2019,2019ApJ...883L..32V}. Our result can be another observational evidence that disagrees with the result of a double loop structure reported in \cite{2008ApJ...689..184M}. 

Meanwhile, \cite{2019ApJ...883L..32V} reported that the stellar stream on the east side appears to pass through NGC 5907 and extends long to the northwest. For direct comparison, we present the zoomed-in images of both ours and \cite{2019ApJ...883L..32V}\footnote{This $g$-filter image is provided from the website \url{https://www.pietervandokkum.com/ngc5907}} with the same FoV (the right panels in Figure \ref{fig:fig6}). We could find no definite evidence of any structure extending to the northwest in our image. This result seems to be reasonable since \cite{2019ApJ...883L..32V} reported that the structure on the west side of the galaxy has a surface brightness fainter than 28 mag arcsec$^{-2}$ in both $g$- and $r$-filters, which is marginally below our surface brightness limit. 

One might argue that there seem to be sparse, faint clumps along the reported long tail. However, it cannot be evidence of the tail because it is difficult to distinguish whether it is a part of a tail structure or a fluctuation caused by unmasked diffuse light from bright sources. Back to the point, the existence of the stellar stream on the east side seems quite certain due to the continuity. On the other hand, in the case of the west side, we would not have been able to claim the existence of the tail at all without comparing it with the result of \cite{2019ApJ...883L..32V}. 

\subsection{Comparison with other studies}

Describing our telescope's performance by comparing the imaging result with those of other surveys may be inappropriate because all imaging surveys have been conducted under different photometric conditions and using various observation strategies such as different exposure times and dithering patterns. In addition, multiple factors in the observation system, such as aperture size, detector type, and filter sets, are also involved with the telescope's performance. Despite this complexity, comparing several characteristics of imaging surveys that have achieved similar results can help us understand our telescope's efficiency in LSB detection. 
%It is almost impossible to objectively state the performance of telescopes because all imaging surveys are conducted under different photometric conditions and using various observation strategies such as different exposure times and dithering patterns. In addition, the performance of telescopes involves multiple factors in the observation system, such as aperture size, detector type, and filter sets. Therefore, it may not be appropriate to describe the performance of our telescope just by comparing the results of various imaging surveys. Nevertheless, comparing several factors can help us understand the efficiency of the telescope in terms of LSB detection. 

\begin{deluxetable*}{lcccc}[t]
\tablenum{1}
\tablecaption{Key characteristics of the studies that observed the stellar stream of NGC 5907 \label{tab:comparison}}
\tablewidth{0pt}
\tablehead{
\colhead{Reference} & \colhead{Diameter} & \colhead{Total Exposure} & \colhead{filters} & \colhead{Telescope} \\
& \colhead{(m)} & \colhead{(hr)} & 
}
%\decimalcolnumbers
\startdata
%\cite{1998ApJ...504L..23S} & 0.6/0.9 & 26.17 & m$_\mathrm{6660}$ & 1.71 & Schmidt telescope \\
%\cite{2008ApJ...689..184M} & 0.5 & 11.4 & $L$ & 0.45 & Ritchey--Chr\'etien telescope \\
%\cite{2016AJ....152...72L} & 8.2 & 0.2 & $gri$ & 0.2 & Subaru telescope \\
%\cite{muller+2019} & 1.4 & 7.2 & $L$ & 0.4 & Milankovi\'c telescope \\
%\cite{2019ApJ...883L..32V} & 0.143 & 4.8 & $gr$ & 2.8 & Dragonfly Telephoto Array \\
%This study & 0.3 & 2.0 & $L$ & 1.89 & K-DRIFT pathfinder \\
\cite{1998ApJ...504L..23S} & 0.6/0.9 & 26.17 & m$_\mathrm{6660}$, m$_\mathrm{8020}$ & Schmidt telescope \\
\cite{2008ApJ...689..184M} & 0.5 & 11.4 & $L$ & Ritchey--Chr\'etien telescope \\
\cite{2016AJ....152...72L} & 8.2 & 0.2 & $g$, $r$, $i$ & Subaru telescope \\
\cite{muller+2019} & 1.4 & 7.2 & $L$ & Milankovi\'c telescope \\
\cite{2019ApJ...883L..32V} & 0.143 (1) & 115 (4.8) & $g$, $r$ & Dragonfly Telephoto Array \\
This study & 0.3 & 2.0 & $L$ & K-DRIFT pathfinder \\
\enddata
\tablecomments{The values presented here are only for a single aperture with a single filter. Therefore, the Dragonfly Telephoto Array can practically be regarded as a 1 m telescope with a total exposure time of 4.8 hr.}
\end{deluxetable*}

Table \ref{tab:comparison} presents the key characteristics of previous observations that identified the stellar stream of NGC 5907. Note that not all the studies have been aimed at LSB research, so their data reduction may not have been optimized. Our telescope has the smallest aperture size\footnote{The Dragonfly Telephoto Array consists of $48 \times 0.143$ m lenses. Its single aperture is smaller than ours, but the array can practically be regarded as a single 1 m telescope.} and moderately short total exposure time but could achieve comparable performance in LSB detection. 

It is worth noting that \cite{2019ApJ...883L..32V} achieved a deeper surface brightness limit than ours, owing to the time-efficient array system. Their observation time was about 2.5 times longer than ours, but the practical total integration time was nearly 60 times longer. According to the principle that background noise is reduced by the square root of the number of co-added images, using our telescope would not require such a long integration time to achieve a result similar to that of \cite{2019ApJ...883L..32V}. It is expected to achieve a sufficient surface brightness limit if a total integration time of only 2--3 times longer than the present. 

Instead, as a telescope with a single aperture, the description comparing characteristics with the Subaru telescope can be more concise. Since \cite{2016AJ....152...72L} did not specify the surface brightness limit from the background noise, let us assume the photometric depth is similar to ours. The aperture size of our telescope is about 30 times smaller than that of the Subaru telescope, while our total exposure time is ten times longer. Due to adopting the luminance filter, our telescope can collect about twice as much light over the same time. This suggests that our telescope is $\sim$$30^2/10/2=50$ times more efficient than the Subaru telescope, at least in some sense, in detecting LSB features. 

%% The "ht!" tells LaTeX to put the figure "here" first, at the "top" next
%% and to override the normal way of calculating a float position
\begin{figure*}[t]
\centering
\vspace{3mm}
\includegraphics[width=150mm]{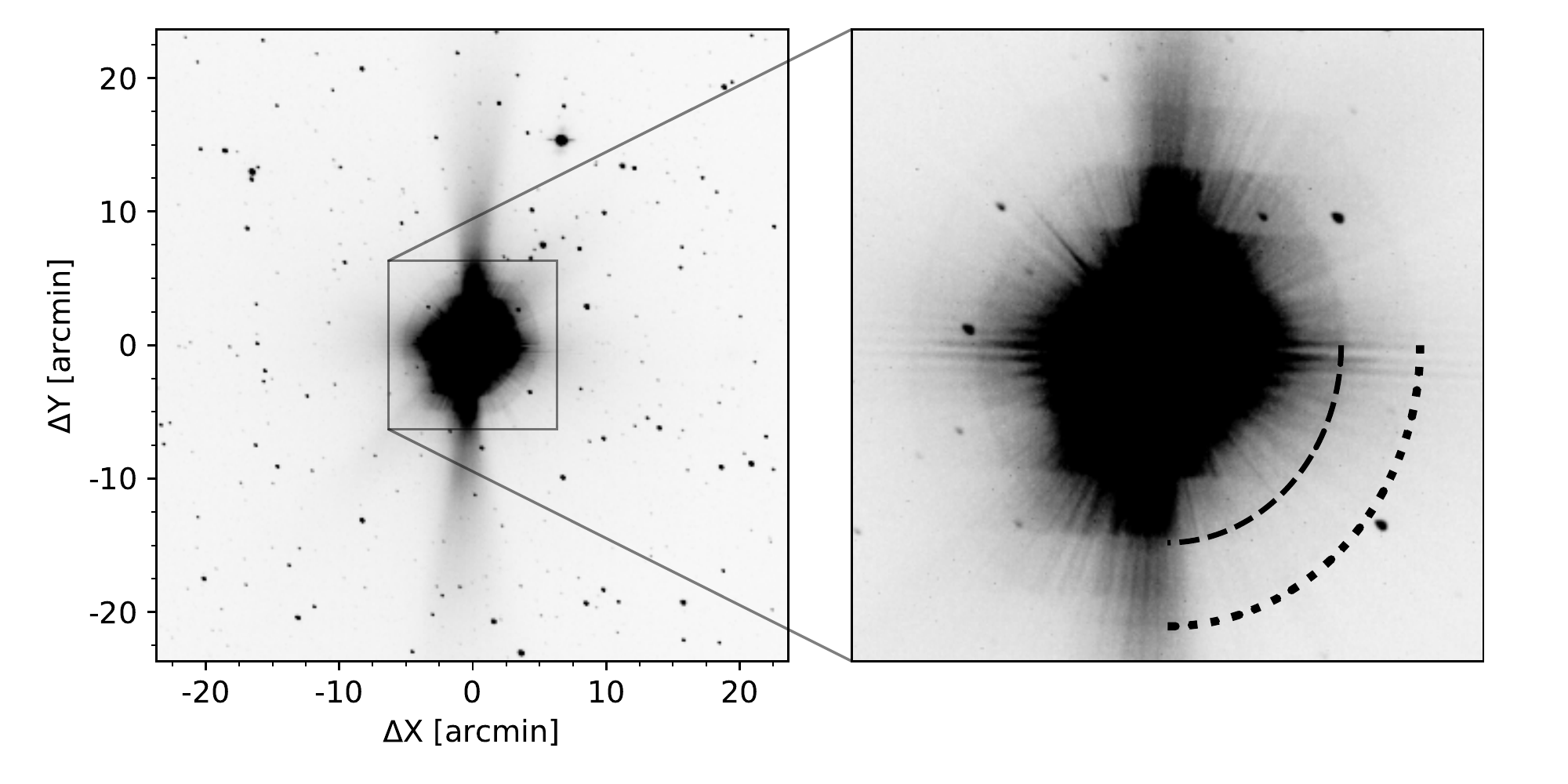}
\caption{Example of a 30-sec exposure of Arcturus. Not only scattered light but also reflection features are clearly shown. The dashed and dotted arcs indicate the reflection edges, which might be created from the dewar window and filter. The radii of annuli are $\sim$4 and $\sim$5.6 arcmin, respectively. \label{fig:fig7}}
\end{figure*}

%% The "ht!" tells LaTeX to put the figure "here" first, at the "top" next
%% and to override the normal way of calculating a float position
\begin{figure*}[t]
\centering
\vspace{3mm}
\includegraphics[width=150mm]{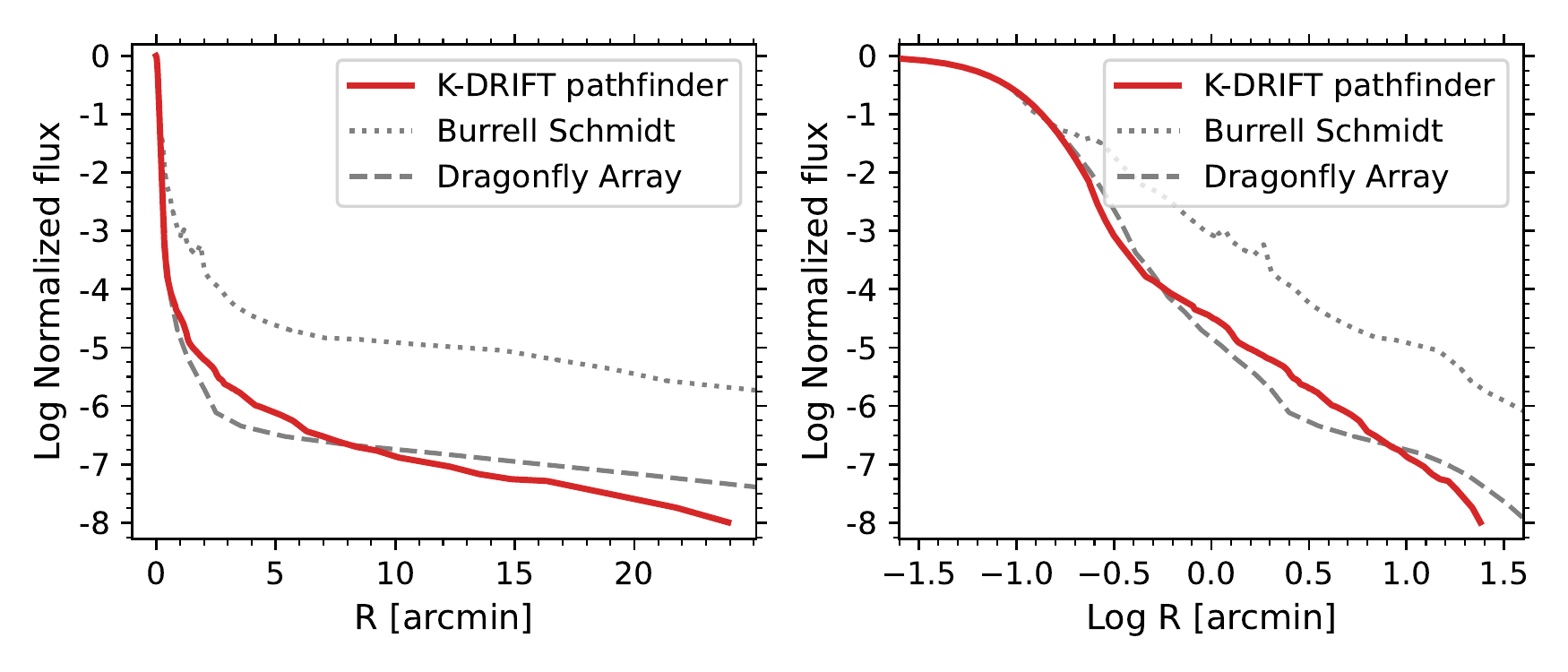}
\caption{Comparison of the PSF of our telescope (red) with those of the Burrell Schmidt Telescope \citep[dotted;][]{2009PASP..121.1267S} and the Dragonfly Telephoto Array \citep[dashed;][]{2014PASP..126...55A}. The fluxes of the three PSF profiles were adjusted so that the fluxes at a radius of 0.1 arcmin were identical. The left and right panels show the PSF profiles as a function of radius on linear and logarithmic scales, respectively. \label{fig:fig8}}
\end{figure*}

\subsection{Scattering and reflection} \label{sec:scatter}

Apart from the deeper surface brightness limit of \cite{2019ApJ...883L..32V}, a significant unmasked light can be seen in the Dragonfly Telephoto Array image, especially in the lower right corner (Figure \ref{fig:fig6}). Even considering the different object-mask sizes, our data seems less affected by scattered light. Thus, in addition to identifying the faint stellar stream of NGC 5907, it would be worthwhile to examine the scattered and reflected light. Indeed, a large PSF can have a significant impact on estimating the brightness and shape of LSB features \citep[see][]{2008MNRAS.388.1521D,2011ApJ...731...89T,2014A&A...567A..97S,2016ApJ...823..123T}. 

The detailed analysis of scattering and reflection is out of the scope of this paper. Instead, we examined the PSF profile of our telescope using a bright star. Before that, it would be helpful to check the appearance of the scattering and reflection visually. Figure \ref{fig:fig7} presents an example of a 30-sec exposure of Arcturus. The most conspicuous artifact is a vertical jet-like feature on a scale of tens of arcmin. It appears to have a slightly misaligned and diffuse shape, unlike the typical saturation bleed. The right panel emphasizes an asterisk-like scattered light and reflected light from the dewar window and filter. The outer annulus was revealed to be out to a radius of 5.6 arcmin. We will discuss these features in the next section. 

The PSF core was parameterized using PSFEx \citep{2011ASPC..442..435B} that extracts a precise model of unsaturated field stars. The extended PSF wing was modeled from Arcturus using a simple circular isophote fitting to avoid the effect of the complex structures. Here, we could construct the profile in the saturated area of the star by connecting the fragmented wing profiles modeled from the images with different exposure times (e.g., 1, 10, 30, and 60 sec). 

Figure \ref{fig:fig8} shows the radial profile of the PSF model of our telescope. The PSF's core size is pretty large (FWHM $\sim$ 6--8 arcsec), but the wing appears very compact and faint. To compare its shape, we also presented the PSF models of the Burrell Schmidt telescope \citep{2009PASP..121.1267S} and the Dragonfly Telephoto Array \citep{2014PASP..126...55A}. The PSF wing reaches a very low level, as low as about 100 times fainter than the former on average. The overall trend of the PSF profile is comparable to the latter but seems slightly different. While the outermost wing of our PSF model is fainter than that of the Dragonfly Telephoto Array, an excess exists between the radii of $\sim$0.5--8 arcmin in our PSF. This seems to be caused by the scattering and reflection of light shown in Figure \ref{fig:fig7}. Nevertheless, we conclude that this is a very encouraging result compared to the PSF level of conventional telescopes. We remind that this result is preliminary because we used only a single bright star centered on the image. Indeed, the characterization of the wide-angle PSF may be sensitive to sky subtraction, particularly in crowded regions \citep[cf.][]{2022ApJ...925..219L}. We will improve the telescope's PSF model analysis in future work. 

%We also compared the PSF model with the surface brightness profile of NGC 5907\footnote{The geometric parameters of the isophote fitting for NGC 5907 were considered free parameters except for the central position. The error budget in the surface brightness was computed by combining the isophote fitting error and the uncertainty of the sky determination.} in the same extension (Figure \ref{fig:fig8}). The surface brightness profile is fitted slightly beyond the surface brightness limit. However, the outside of the radius of $\sim$6 arcmin seems not a real structure but rather a background affected by the PSF \citep[see also][]{2022arXiv220406596G}. This suggests that a deeper surface brightness limit and additional control of the PSF effect are needed to investigate the stellar halo, which is more challenging to study due to its faintness and diffuseness. 

\section{Discussion} \label{sec:disc}

Although this study achieved rather encouraging results in LSB detection, several unexpected issues were revealed in the detector and the mirror. This section introduces and discusses the issues that must be addressed in future work to improve data quality and photometric depth. 

First, our data seems to be affected by false signals, namely the RBI.\footnote{In general, the RBI effect is known to mainly occur in front-illuminated CCD chips. The cause is unknown, but the same phenomenon occurred in this study with a back-illuminated sCMOS chip.} In particular, the RBI effect can be clearly seen in the dark frame taken after object exposure, specifically as follows: (1) the dark level was slightly raised by the high background level, and (2) faint spots were left where bright stars were placed on the last object frame. These traces gradually faded and disappeared after at least $\sim$10 min. This means that the RBI effect may affect almost all science images taken at 30-sec intervals. However, the impact on the background fluctuation in the final co-added image is not expected to be very significant since the RBI level is only $\sim$0.5\% of the original signal and disappears during dithering observation. 

Nevertheless, the RBI effect must be eliminated. If the detector is the cause, it can be solved by replacing it with another detector. Otherwise, we must find other ways to minimize this effect. The RBI effect is known to be temperature-dependent. The lower the chip's temperature, the longer trapped electrons can remain. In addition, the brighter the object, the stronger the residual in a shorter time. In future work, we will characterize the RBI effect of the CMOS chip depending on temperature and exposure time so that we can determine an optimized observation strategy. 

Second, the PSF's core size is twice as large as expected from the optics design. Even considering the average seeing of the observatory ($\sim$2--3 arcsec), it must be smaller than the present. Some test observations not introduced in this paper suggest that the large PSF is unlikely due to focusing or tracking problems. In addition to this, the (vertically) scattered light was clearly seen around bright stars. Ideally, this problem should not occur because our telescope was designed not to scatter any light on the optical path by removing internal structures. We suspect that the secondary mirror with insufficient surface precision mainly causes these problems based on optics simulation using mirror models provided through surface measurement. The secondary mirror will be replaced by a new one manufactured more precisely in mid-2022, and then the problem is expected to be solved or at least reduced. Another possibility is that it might be caused by diffraction from the microlenses on the CMOS array. We will also check this issue after replacing the secondary mirror. 

Lastly, internal reflections by the dewer window and filter mimic an extended halo and contaminate the vicinity of bright stars. This problem can be reduced by using anti-reflective coatings on the filter and dewer window. However, faint reflections may still appear in practice. Eventually, a software solution must be invented to eliminate these features. The reflection features may have diverse brightness and central position depending on the position and brightness of the object on the image. Therefore, we will characterize the reflection through photon simulation with the exact specifications of our telescope in future work. 

\section{Summary} \label{sec:sum}

A simple optics design is advantageous for studying LSB nature because it can reduce photometric uncertainty arising from the instrument, thereby achieving extremely deep surface brightness limits. For this reason, we have been developing new telescopes adopting an optics design with a clean pupil to observe LSB features effectively. In early 2021, we successfully manufactured a prototype telescope called the K-DRIFT pathfinder. This telescope has an off-axis design consisting of a linear astigmatism free-three mirror system. Since there is no structure on the optical path, it was expected to minimize the loss and scattering of light. 

To assess the telescope's performance, we observed NGC 5907, which is well known to have a large stellar stream. The data taken for two days were processed through the optimized data reduction, including flat fielding using a dark-sky flat and dedicated sky subtraction. We obtained a wide and deep image with a total integration time of 2 hr by co-adding reduced images. As a result, we were able to achieve a 1$\sigma$ surface brightness limit of $\sim$28.5 mag arcsec$^{-2}$ in $10^{\prime\prime}\times10^{\prime\prime}$ boxes and identify a single curved stellar stream to the east of NGC 5907. Compared to other observational studies that have shown similar detection ability, one can say that our telescope is more efficient in detecting LSB features at face value. In addition, we found that the PSF wing reached a very low level despite being affected by some reflection features. 

We highlight that this is an encouraging result, considering the relatively small aperture and short integration time. Data with similar conditions to this study would help detect tidal features and faint dwarf galaxies. However, it seems insufficient to identify fainter structures such as stellar halos in galaxies because of much lower surface brightness levels and their diffuseness. We expect this limitation to be overcome by increasing exposure time and applying more effective observation strategies that allow the background to be as flat as possible, such as adding rotation to the dithering pattern. 

We note that the K-DRIFT pathfinder is a prototype of four to five telescopes to be built over the next decade. Through this study, we understood what to upgrade to improve the performance of the next K-DRIFT telescopes. For example, more precise mirror manufacturing is required, and the detector's characteristics, including the RBI effect, must be clarified. At the same time, data must be expanded through intensive surveys for scientific achievements. By adopting a larger FoV and various filter sets, we expect to detect LSB features more effectively and analyze the LSB nature in detail. 

%% IMPORTANT! The old "\acknowledgment" command has been depreciated. It was
%% compile with \acknowledgment, you will get an error print to the screen

\begin{acknowledgments}
We are grateful to an anonymous referee for constructive comments and suggestions.
This research was supported by the Korea Astronomy and Space Science Institute under the R\&D program (Project No. 2022-1-830-05), supervised by the Ministry of Science and ICT. This research was supported by `National Research Council of Science \& Technology (NST)' - `Korea Astronomy and Space Science (KASI)' Postdoctoral Fellowship Program for Young Scientists at KASI in South Korea.

\end{acknowledgments}

%% in the research for the paper.  Each keyword is checked against the master 
%% parentheses after the keyword, but they are not verified.

\vspace{5mm}
\facilities{The K-DRIFT pathfinder, SDSS, IRSA, NED.}


\software{Astropy \citep{astropy:2013,astropy:2018}, Scipy \citep{2020SciPy-NMeth}, Source Extractor \citep{1996A&AS..117..393B}, SWarp \citep{2002ASPC..281..228B}, Photutils \citep{2021zndo...4624996B}, Matplotlib \citep{Hunter:2007}, PSFEx \citep{2011ASPC..442..435B}}



%\appendix

%% For this sample, we use BibTeX plus aasjournals.bst to generate the

\bibliography{ms}{}
\bibliographystyle{aasjournal}



\end{document}